\documentclass[a4paper,11pt]{article}
\pdfoutput=1 % if your are submitting a pdflatex (i.e. if you have
\usepackage{jcappub} % for details on the use of the package, please
\usepackage[T1]{fontenc} % if needed

\title{\boldmath A title with some math: $x=1$}

\usepackage{graphicx}% Include figure files
\usepackage{dcolumn}% Align table columns on decimal point
\usepackage{amssymb}% outlined letters
\usepackage{amsmath}
\usepackage{color}
\usepackage{verbatim}
\usepackage{multirow}
\usepackage{soul}

\def\la{\lambda}

\def\ta{\tau}

\def\La{\Lambda}

\newcommand{\ben}{\begin{equation}}
\newcommand{\een}{\end{equation}}
\newcommand{\bea}{\begin{eqnarray}}
\newcommand{\eea}{\end{eqnarray}}
\newcommand{\ba}{\begin{array}}
\newcommand{\ea}{\end{array}}
\newcommand{\bit}{\begin{itemize}}
\newcommand{\eit}{\end{itemize}}

\newcommand{\bk}{\textbf{k}}

\newcommand{\fd}{f_{10}}

\newcommand{\Obhh}{\Omega_{\mathrm{b}}h^{2}}

\newcommand{\Ol}{\Omega_{\Lambda}}

\newcommand{\CMBEASY}{\textsc{CMBeasy }}

\newcommand{\PL}{$\mathcal{PL}$}
\newcommand{\PLr}{$\mathcal{PL}+r$}
\newcommand{\PLgmu}{$\mathcal{PL}+G\mu$}

\newcommand{\PLrgmu}{$\mathcal{PL}+r+G\mu$}

\newcommand{\tEq}{\tau_\mathrm{eq}}
\newcommand{\tLam}{\tau_\mathrm{\La}}
\newcommand{\tNow}{\tau_0}

\newcommand{\Planck}{{\it Planck}}
\newcommand{\LCDM}{$\Lambda$CDM}

\newcommand{\ie}{\textit{i.e.\ }}

\newcommand{\tInit}{\ta_\text{i}}

\newcommand{\eVec}{c}
\newcommand{\sFun}{s}
\newcommand{\eVal}{\la}

\title{New CMB constraints for Abelian Higgs cosmic strings}

\newcommand{\addressSussex}{Department of Physics \& Astronomy, University of Sussex, \\Brighton, BN1 9QH, United Kingdom}
\newcommand{\HIPetc}{\affiliation[e]{
		Department of Physics and Helsinki Institute of Physics,
		\\PL 64,  
		FI-00014 University of Helsinki,
		Finland
	}}

\author[a]{Joanes Lizarraga,}
\emailAdd{joanes.lizarraga@ehu.eus}
\affiliation[a]{Department of Theoretical Physics, University of the Basque Country UPV-EHU, \\48040 Bilbao, Spain}

\author[a]{Jon Urrestilla,}
\emailAdd{jon.urrestilla@ehu.eus}

\author[b,c]{David Daverio,}
\emailAdd{david.daverio@unige.ch}
\affiliation[b]{African Institute for Mathematical Sciences, \\6 Melrose Rd, Muizenberg, 7945, Cape Town, South Africa}
\affiliation[c]{D\'epartement de Physique Th\'eorique and Center for Astroparticle Physics, \\Universit\'e de Gen\`eve, \\24 quai Ansermet, CH--1211 Gen\`eve 4, Switzerland}

\author[d,e]{Mark Hindmarsh} 
\emailAdd{m.b.hindmarsh@sussex.ac.uk}
\affiliation[d]{\addressSussex}
\HIPetc

\author[c]{and Martin Kunz}
\emailAdd{martin.kunz@unige.ch}

\abstract{We present cosmic microwave background (CMB) power spectra from recent numerical simulations of cosmic strings in the Abelian Higgs model and compare them to CMB power spectra measured by \Planck. We obtain revised constraints on the cosmic string tension parameter $G\mu$. For example, in the \LCDM\ model with the addition of strings and no primordial tensor perturbations, we find $G\mu < 2.0 \times 10^{-7}$ at 95\% confidence, about 20\% lower than the value obtained from previous simulations, which had 1/64 of the spatial volume. The increased computational volume also makes it possible to simulate fully the physical equations of motion, in which the string cores shrink in comoving coordinates.  We find however that this, and  the larger dynamic range, changes the amplitude of the power spectra by only about 10\%. The main cause of the stronger constraints on $G\mu$ is instead an improved treatment of the string evolution across the radiation-matter transition.}

\begin{document}
\maketitle

%=====================================================================
%=====================================================================
%=====================================================================

\section{Introduction}
\label{sec:intro}

The Cosmic Microwave Background (CMB) has demonstrated many times its power as a clean and reliable probe for early and late-time cosmological physics. The latest results from the \Planck\ collaboration \cite{Ade:2015xua} have reached percent-level precision on nearly all standard parameters, in some
cases doing even better. The CMB is also an important probe for cosmic defects \cite{Kibble:1976sj} as it is mainly sensitive to the large-scale properties of the defect distribution which is relatively well understood \cite{Pen:1997ae,Pogosian:1999np,Martins:2000cs,Durrer:2001cg,Martins:2003vd,Bevis:2010gj,BlancoPillado:2011dq}. The \Planck\ data currently limits the contribution of strings and other defects to the temperature anisotropies on large scales to be of the order of a few percent or less \cite{Ade:2013xla,Lizarraga:2014xza,Lazanu:2014eya,Ade:2015xua,Charnock:2016nzm}.

However, the limits can be only as accurate and precise as the theoretical forecasts that are compared to the data. In this paper we use the recently updated unequal-time correlation functions (UETC) from \cite{Daverio:2015nva} to compute new CMB power spectra for cosmic strings in the Abelian Higgs  model, the prototypical field theory with such topological defects.   \footnote{For reviews of cosmic strings in cosmology and a discussion of the differences between calculations based on field theory and modelling with ideal Nambu-Goto strings see Refs.~\cite{Copeland:2011dx,Hindmarsh:2011qj}.} The new UETCs are based on simulations that are four times larger (64 times larger in volume) than those used previously \cite{Bevis:2010gj}, \ie\ they have a four times larger dynamical range in both space and time. As usual, the Abelian Higgs (AH) strings were simulated with couplings at the TypeI/TypeII boundary.

Thanks to the much larger simulation volume, we were able to simulate directly scales that previously we could only infer. It was also possible to keep the string cores at a constant physical width, instead of letting them grow with the expansion of the universe as was done previously. In fact, in previous works with smaller simulation volumes, the string width was allowed to grow artificially in order to keep the core of the string resolved in a grid in comoving coordinates \cite{Press:1989yh,Bevis:2010gj}. The growth of the string was controlled by a parameter called $s$, which ranges from 0 to 1: $s=0$ corresponds to a string whose width stays constant in comoving coordinates, and $s=1$ corresponds to strings without artificial core-growth, \ie\ with a constant physical width throughout the simulation. In the new simulations presented in \cite{Daverio:2015nva}, we were able to simulate strings with $s=1$. In addition, the modelling of the evolution of the string network across the cosmological radiation-matter and matter-$\Lambda$ transitions has been treated much more carefully than in previous works.

The differences in observables between between $s=1$ and $s=0$ were not great, and in particular the equal time correlators of the new simulations were consistent with the old ones (see Fig. 6 of ref.~ \cite{Daverio:2015nva}). The consistency meant that we were able to merge the UETCs from $s=1$ simulations with ones computed at $s=0$, thereby increasing the accessible range of time differences. A more detailed comparison of $s=1$ and $s=0$, and tests of the scaling assumption, will be the subject of a future publication.

In the next section we describe briefly how we compute the power spectra of temperature and polarisation anisotropies in the CMB from the new UETCs obtained in  \cite{Daverio:2015nva}. In section \ref{sec_cmbLAH} we compare how the different improvements in the simulations affect the $C_\ell$ and what the main causes of the differences with the previous spectra are. In particular we demonstrate that the new simulations change the $C_\ell$ by only about 10\%, when they are computed with the same methods as in \cite{Bevis:2010gj}. We then use the new $C_\ell$ to place constraints on the abundance of cosmic strings in section \ref{sec:fits} before concluding. 

%=====================================================================
%=====================================================================
%=====================================================================

\section{Method overview}
\label{sec:method}

The UETC method of calculating power spectra \cite{Pen:1997ae,Durrer:2001cg} exploits the observation that the unequal time correlators of the energy-momentum tensor contain all the required information. UETCs are defined as follows:

\begin{equation}
{U}_{\lambda\kappa\mu\nu} (\textbf{k},\tau,\tau') = \langle{\mathcal{T}}_{\lambda\kappa}(\textbf{k},\tau){\mathcal{T}}^*_{\mu\nu}(\textbf{k},\tau')\rangle\,,
\label{emtens}
\end{equation}
where ${\mathcal{T}}_{\alpha\beta}(\textbf{k},\tau)$ is the energy-momentum tensor of the field theory, in this case the Abelian Higgs model.

Due to rotational symmetry, we need only calculate correlations between four projected components of the energy-momentum tensor
\ben
S_a(\bk,t) = P_a^{\mu\nu}(\bk) {\mathcal{T}}_{\mu\nu}(\textbf{k},\tau'),
\een
where $P_a^{\mu\nu}(\bk)$ project onto scalar, vector and tensor parts.  In principle there are two of each, but the two vector and the two tensor components are related by parity. Hence, we may consider that the indices $a$, $b$ take four values corresponding to the independent components of the energy momentum tensor: two scalar, one vector and one tensor. We will denote the scalar indices $1$ and $2$ (corresponding to the longitudinal gauge potentials $\phi$ and $\psi$), the vector component with `v' and the tensor component with `t'. 

Thus we can write 
\begin{equation}
{U}_{ab}(\textbf{k} ,\tau,\tau') = \frac{\phi_0^4}{\sqrt{\tau\tau'}}\frac{1}{V}{C}_{ab}(k,\tau,\tau'),
\label{UETCdecom}
\end{equation}
where $\phi_0$ is the symmetry breaking scale, $V$ a formal comoving volume factor, and the functions $C_{ab}(k,\tau,\tau')$ defined by this equation are dimensionless. Note that the scalar, vector and tensor contributions are decoupled for linearized cosmological perturbations, and therefore cross correlators between them vanish, except in the scalar sector: hence the 5 independent correlators. 

The UETCs give the power spectra of cosmological perturbations when convolved with the appropriate Green's functions. This is accomplished by decomposing them into a set of functions derived from the eigenvectors of the UETCs, used as sources for an Einstein-Boltzmann integrator. The power spectrum of interest is reconstructed as the sum of power spectra from each of the source functions . 

When the times  $\tau$ and $\tau'$ are both in same cosmological epoch, the correlation functions depend only on the dimensionless combinations $x = k\tau$ and $x' = k\tau'$. This behaviour is called scaling. A scaling UETC will have eigenvectors which depend on $k$ and $\tau$ only through the combination $x$. It is also convenient to write correlators as functions of $z = \sqrt{xx'}$ and $r = x'/x$.

Perfect scaling is not a feature of the true UETCs, as the universe undergoes a transition from radiation-dominated to matter-dominated expansion during times of interest, and more recently to accelerated expansion with an effective cosmological constant $\La$.  Hence the UETCs also depend explicitly on dimensionful quantities $\tEq$ and $\tLam$, the conformal times of equal radiation and matter density, and equal matter and dark energy density. 

In principle one should calculate the UETCs starting at a time well before horizon entry for the shortest wavelength mode of interest, and simulate until today. In order to obtain a CMB power spectrum suitable for comparison to the \Planck\ data, one would have to achieve scaling over a conformal time ratio of a few thousand, well beyond current capabilities. Instead, we simulate over shorter time intervals, and patch together the results, either at the level of the eigenvectors \cite{Pen:1997ae,Durrer:1998rw,Bevis:2010gj}, or at the level of the UETCs \cite{Daverio:2015nva}, or a combination of both \cite{Fenu:2013tea}.
 
To insert the UETCs as a source of perturbations into an Einstein-Boltzmann (EB) solver we need to take a ``square root" of the UETCs as the EB solvers evolve the perturbation variables, not their power spectra. To this end the UETCs, which are real and symmetric, can be decomposed into their eigenfunctions $\eVec^n(k,\ta)$ defined through 
\ben
\int_{\tInit}^{\tNow} d\ta' C_{ab}(k,\ta,\ta') \eVec_b^n(k,\ta') = \eVal_n(k)\eVec_a^n(k,\ta).
\een
The power spectra and cross-correlations of a perturbation in a cosmological variable $X_a$ can be written \begin{equation}
\langle{X_a}(\textbf{k},\tau){X_b}^*(\textbf{k},\tau)\rangle = \frac{\phi_0^4}{V}\sum_n \la_n I_a^n(k,\tau) I_b^{n*}(k,\tau)\,,
\end{equation}
where the contribution of each linear term, $I_a^n(k,\ta)$, is
\begin{equation}
I_a^n(k,\ta) = \int_{{\tInit}}^{\ta} d\ta' \mathcal{G}^X_{ab}(k,\ta,\ta')\frac{\eVec_b^n(k,\ta')}{\sqrt{\ta'}}\,.
\label{e:EBInt}
\end{equation}
Here,  $\mathcal{G}^X$ is the Green's function for the quantity $X$, which is implemented as a numerical integration by an Einstein-Boltzmann  solver.
The power spectrum can then be constructed as the superposition of the results of integrating the source function
\ben
\sFun_a^n(k,\tau) =  \sqrt{\eVal_n(k)}\eVec_a^n(k,\tau) \, .
\een

As mentioned above, there are several different ways to build the source functions based on the simulation outputs. In \cite{Daverio:2015nva} we developed a new method of combining the information from simulations in the different cosmological epochs (radiation, matter, $\Lambda$) to produce an accurate model of the true UETC at any required value of comoving wavenumber $k$. We found that it improved on the method of Ref.~\cite{Fenu:2013tea}, which in turn was a significant improvement on methods used in our previous work \cite{Bevis:2010gj}. The methods are described in detail in \cite{Daverio:2015nva}. Here we merely summarise them.

%=====================================================================

\subsection{Simple eigenvector decomposition}
\label{subsec:eigen}

In \cite{Bevis:2010gj} we interpolated directly the source functions with a simple interpolation rule,
\ben
\sFun_n(k,\tau) = e_{\Lambda}(\tau) \left( e(\tau) \sFun^\text{R}_n(x) + 
(1 - e(\tau)) \sFun^\text{M}_n(x) \right).
\label{neil_interp}
\een
Note that the eigenfunctions at $\Lambda$ domination are implicitly assumed to be zero. The eigenvector interpolation functions $e(\tau)$ and $e_{\Lambda}(\tau)$ for radiation-matter and matter-$\Lambda$ transitions respectively were taken to be \cite{Bevis:2006mj,Bevis:2010gj},  
\begin{equation}
e(\tau)=\frac{1}{1+\chi[a(\tau)]}\,,
\label{e_Neil}
\end{equation}
\begin{equation}
e_{\Lambda}(\tau)=\frac{1}{1+\chi_{\Lambda}[a(\tau)]}\,,
\label{e_NeilLambda}
\end{equation}
where $\chi[a] = a \Omega_{\mathrm{m}}/\Omega_{\mathrm{r}}$ and $\chi_{\Lambda}[a] = a^3 \Omega_{\Lambda}/\Omega_{\mathrm{m}}$ are the ratio between the density fractions at the given value of the scale factor.

%=====================================================================

\subsection{Multi-stage eigenvector decomposition}
\label{subsec:multistage}

A set of $N_\text{U}$ UETCs at different times $\tau_i$ are defined as interpolations between the radiation and matter era correlators,\footnote{For ease in notation, we will drop the subindices $a,b$ that distinguish between correlators (see e.g.\ Eq.~\ref{UETCdecom}). The procedure described here is followed for all correlators in the same manner.}
\begin{equation}
{C_i} (k,\tau,\tau') = f_i {C}^{\rm R}(k\tau,k\tau') + (1-f_i) {C}^{\rm M}(k\tau,k\tau')\,,
\label{TransitionUETC}
\end{equation}
where $f_i$ decreases from $f_0=1$ to $f_{N_\text{U}}=0$.

The values of $f_i$ are given by a continuous function $f_i = f(\ta_i)$, which is chosen so that the equal time correlators during the transition era are reproduced.
The source functions are then defined from 
\ben
\sFun^n(k,\tau) = \sum_{i=0}^{N_\text{U}} J_i(\tau) \sFun_i^n(x),
\label{f:eVecOld}
\een
where  $J_i(\tau)$ are indicator functions which turn on the $i$th source function between times 
$\ta_i$ and $\ta_{i+1}$. 

In \cite{Daverio:2015nva} it was found that a single function
\begin{equation}
f(\tau)=\left(1+0.24\frac{\tau}{\tau_{\mathrm{eq}}}\right)^{-0.99}\,,
\label{ft_LAH}
\end{equation}
gave the best fit to the ETCs interpolating between the radiation and matter eras.

%=====================================================================

\subsection{Fixed-$k$ UETC interpolation}
\label{subsec:fixedk}
In this method, we note that the global UETC for a given fixed $k$ is a symmetric function of $\tau$ and $\tau'$. Thus, instead of interpolating the eigenvectors (as in Section \ref{subsec:eigen}) or UETCs in time (as in the previous case \ref{subsec:multistage}), we interpolate them in $k$. For each fixed value of $k$, the non-scaling UETC is assembled from the required relative contribution of the scaling UETCs in radiation and matter, determined by the ratios $\tau/\tEq$ and $\tau'/\tEq$. A reasonably accurate form for the interpolation is
\begin{equation}
C(k,\ta,\ta') = f\left(\frac{\sqrt{\ta\ta'}}{\tEq}\right) C^\text{M}(k\ta,k\ta') + \left(1 - f\left(\frac{\sqrt{\ta\ta'}}{\tEq}\right)\right)  C^\text{R}(k\ta,k\ta').
\label{f:UETCmodel}
\end{equation}

The source functions are derived directly from the eigenvalues and eigenfunctions of this global UETC at each value of $k$ required. The transition from matter to $\Lambda$ domination can be added analogously, for more details see \cite{Daverio:2015nva}.

%=====================================================================
%=====================================================================
%=====================================================================

\section{New power spectra and error assessment}
\label{sec_cmbLAH}

Once the source functions have been defined, they can be inserted into a source enabled EB solver to compute the contributions to CMB power spectra due to the presence of topological defects, in our case AH cosmic strings. For this purpose we use a source enabled version of \CMBEASY\ \cite{Doran:2003sy}. The code has been additionally modified to handle source functions of the form of Eq.~(\ref{f:eVecOld}), where the indicator functions point to the corresponding eigenvector slot depending on the time interval (multi-stage) or on the scale $k$ being integrated (fixed-$k$ interpolation method).

The cosmological parameters used for these calculations are the best-fit values obtained by the \Planck\ collaboration \cite{Ade:2015xua}: $h=0.6726$, $\Obhh=0.02225$, $\Ol=0.6844$ and reionization optical depth $\tau_{\rm re}=0.079$. After diagonalisation, the total contribution of strings to temperature and polarization anisotropies is calculated summing the contribution of each individual source function. We observe that convergence of $1\%$ is obtained for $\sim200$ eigenvectors and so as to avoid uncertainties we set the standard number of eigenvectors to $256$.

This section contains the final power spectra and their corresponding error assessments obtained bootstrapping over realizations in the merging process. We classify the different spectra analyzed in the following sections as:
\begin{itemize}
	\item {Case 0:} spectra from \cite{Bevis:2010gj}, where transitions were modeled by simple eigenvector interpolation, as explained in section~\ref{subsec:eigen}.
	\item {Case 1:} new UETCs but using the same simple eigenvector interpolation at the transitions as in case 0.
	\item {Case 2:} new UETCs using multi-stage eigenvector interpolation (section~\ref{subsec:multistage}).
	\item {Case 2.1:} case 2 plus transition from Matter domination to $\Lambda$ domination.
	\item {Case 3:} new UETCs using fixed-$k$ UETC interpolation for both cosmological transitions (section~\ref{subsec:fixedk}).
\end{itemize}

We separately analyze and compare these cases to isolate the changes due to the innovations proposed in \cite{Daverio:2015nva}. In Sec.~\ref{subsec_A} we study the differences introduced by the new UETCs (case 1), with the previous CMB predictions from \cite{Bevis:2010gj} (case 0). The new UETCs  are based on larger simulations that include runs based on the real   ($s=1$) equations of motion. The rest of the subsections study the effects of the different type of interpolations between the different cosmological eras. Secs.~\ref{subsec_B} and \ref{subsec_D} are devoted to the analysis of the effects produced by the new interpolating functions in the context of the multi-stage eigenvector interpolation method (case 2) and the fixed-$k$ UETC interpolation method respectively (case 3). The effects produced by the matter-$\Lambda$ transitions are also studied separately in Sec.~\ref{subsec_C}. 

%=====================================================================

\subsection{Case 0 $\rightarrow$ 1: effects of the new UETCs}
\label{subsec_A}

\begin{figure}[h!]
\centering
\includegraphics[width=0.49\textwidth]{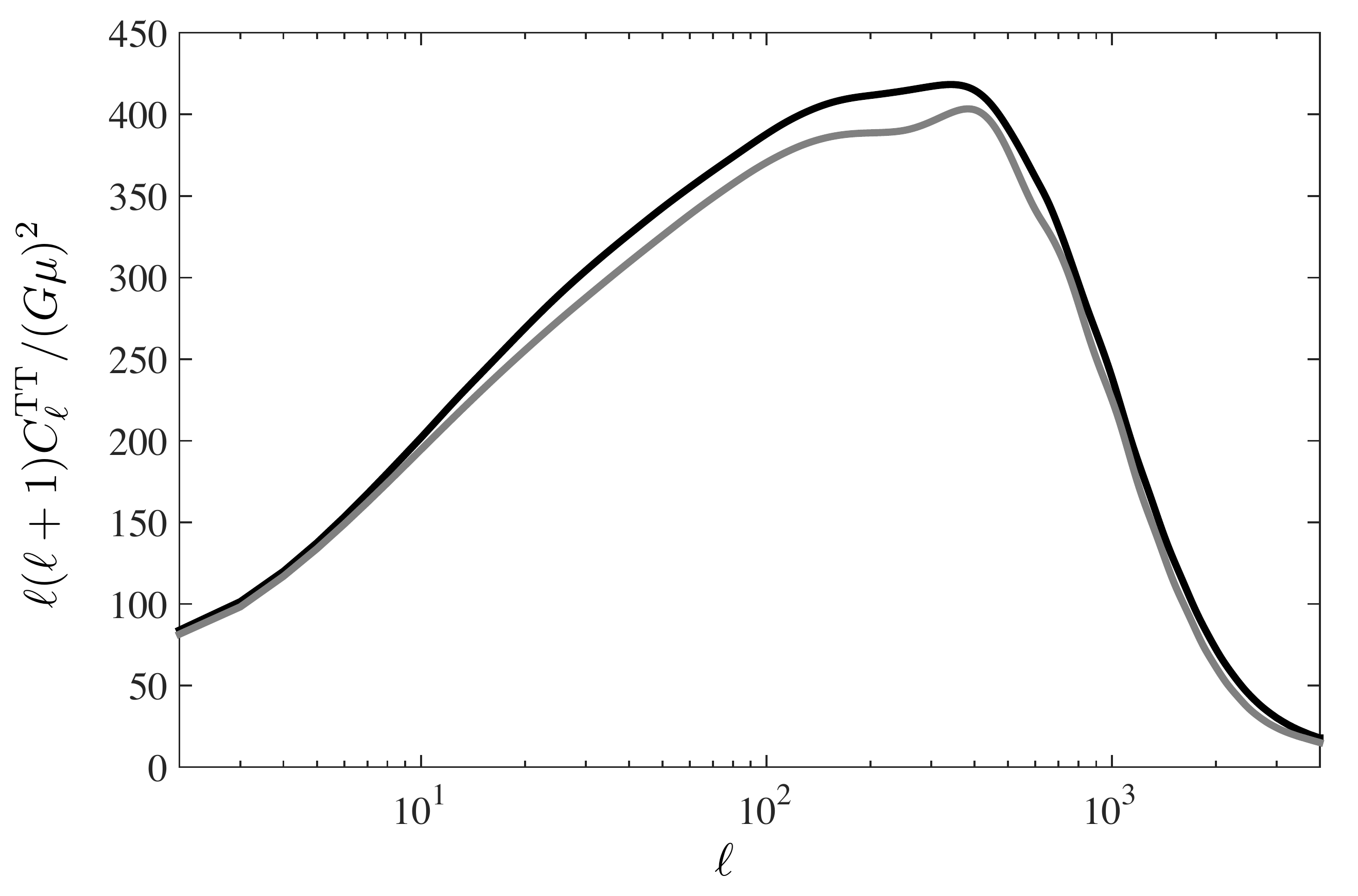}
\includegraphics[width=0.49\textwidth]{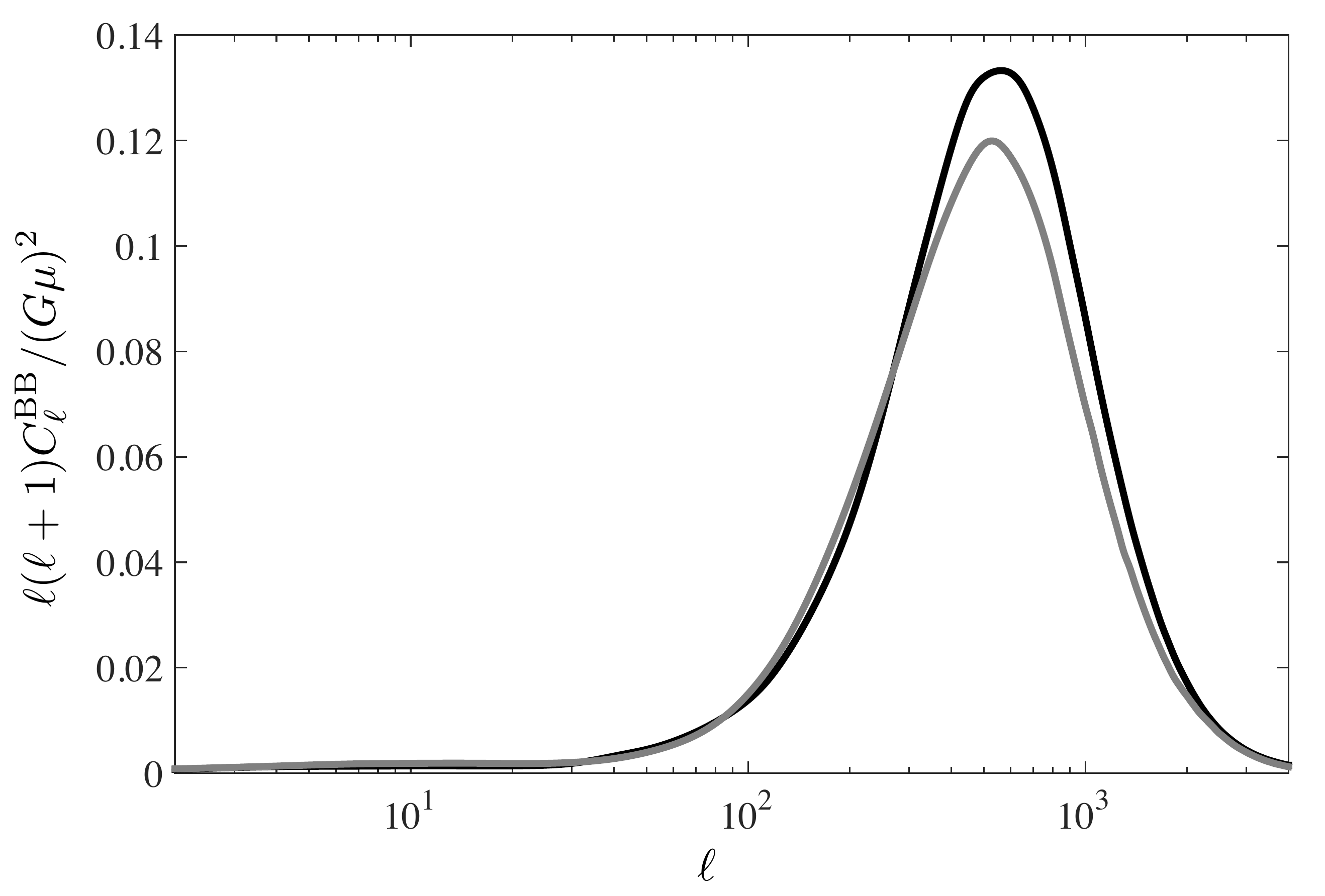}
\caption{Comparison of the spectra obtained in \cite{Bevis:2010gj} (grey line) and in this work (black line). The latter are obtained using UETCs from the new simulations described in \cite{Daverio:2015nva} but using the same simple eigenvector interpolation method as \cite{Bevis:2010gj}. The difference gives an indication of the change coming from the improved simulation of AH strings using a larger simulation volume and the true equations of motion.}
\label{figure_cl1}
\end{figure}

In this section we study the impact of the new simulations presented in \cite{Daverio:2015nva} relative to the previous CMB predictions from \cite{Bevis:2010gj} when using the same methodology to compute the CMB spectra. The new source functions have been calculated from the largest field theory simulations of AH cosmic strings to date, covering a spatial volume 64 times bigger than in \cite{Bevis:2010gj}. We thus have a four times larger dynamical range both in space and in time, and scales that previously could only be inferred by extrapolation can now be directly simulated. The new simulations measure more accurately horizon scale correlations and the small scale power-law behavior. Moreover, it has been possible to reproduce the real equations of motion, \ie $s=1$, and reach the scaling regime both in matter and radiation domination eras. All these improvements are taken into account in the new merged UETC functions.

In order to capture the effect created by the new UETCs, the spectra presented in this section have been calculated using the same simple eigenvector interpolation method used in \cite{Bevis:2010gj}, where the transition is driven by a density weighted interpolation function (see Eq.~(\ref{neil_interp}), (\ref{e_Neil}) and (\ref{e_NeilLambda})).

Figure~\ref{figure_cl1} shows the comparison of the temperature and B-modes for this case (the changes in the EE and TE channels are of similar magnitude). The spectra corresponding to the 2010 paper ({case 0}) are in grey, whereas the new ones ({case 1}) in black. As it can be seen the difference is not very significant. The new spectra are smoother and the small details around the peak of the temperature spectrum have disappeared. The amplitude increases by about 10\%. With such small deviations, the UETCs of \cite{Bevis:2010gj} are sufficient at the level of accuracy claimed in that paper, \ie this result shows that the approximations and extrapolations in \cite{Bevis:2010gj} were justified.

%=====================================================================

\subsection{Case 1 $\rightarrow$ 2: Effects of the new radiation-matter transition function and multi-stage method}
\label{subsec_B}

In this section we analyze the effects produced by the multi-stage cosmological transition treatment and the new interpolation function of Eq.~(\ref{ft_LAH}). In order to focus only on the effects of the transition, all spectra are calculated using the new UETCs.

The multi-stage interpolation method was proposed in \cite{Fenu:2013tea} and it interpolates between eigenvectors of UETCs calculated at intermediate time intervals between radiation and matter, see Eq.~(\ref{TransitionUETC}). The most important difference is that, while in the simple eigenvector interpolation method only the eigenvector of the pure radiation and matter correlators are considered, the multi-stage method derives an orthonormal set of eigenvectors at each desired time slicing. In this interpolation scheme, we observe that $N_{\rm U}=11$ time slices are enough to reach convergence.

In addition, the interpolating function that performs the transition from radiation domination to matter domination has been substituted by the ETC weighted interpolation function of Eq.~(\ref{ft_LAH}). The matter-$\Lambda$ transition will be similarly analyzed in the next section.

\begin{figure}[h!]
\centering
\includegraphics[width=0.49\textwidth]{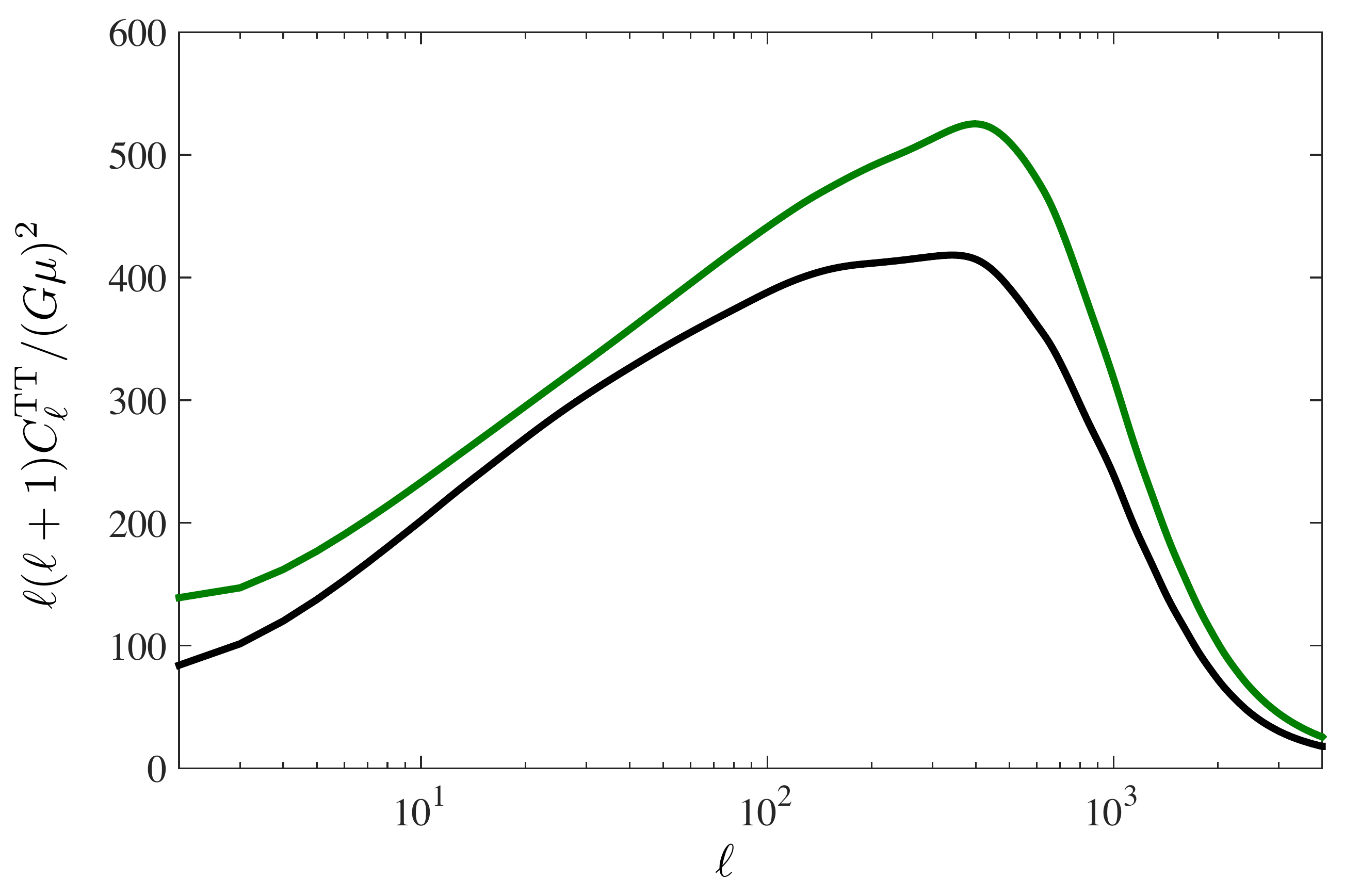}
\includegraphics[width=0.49\textwidth]{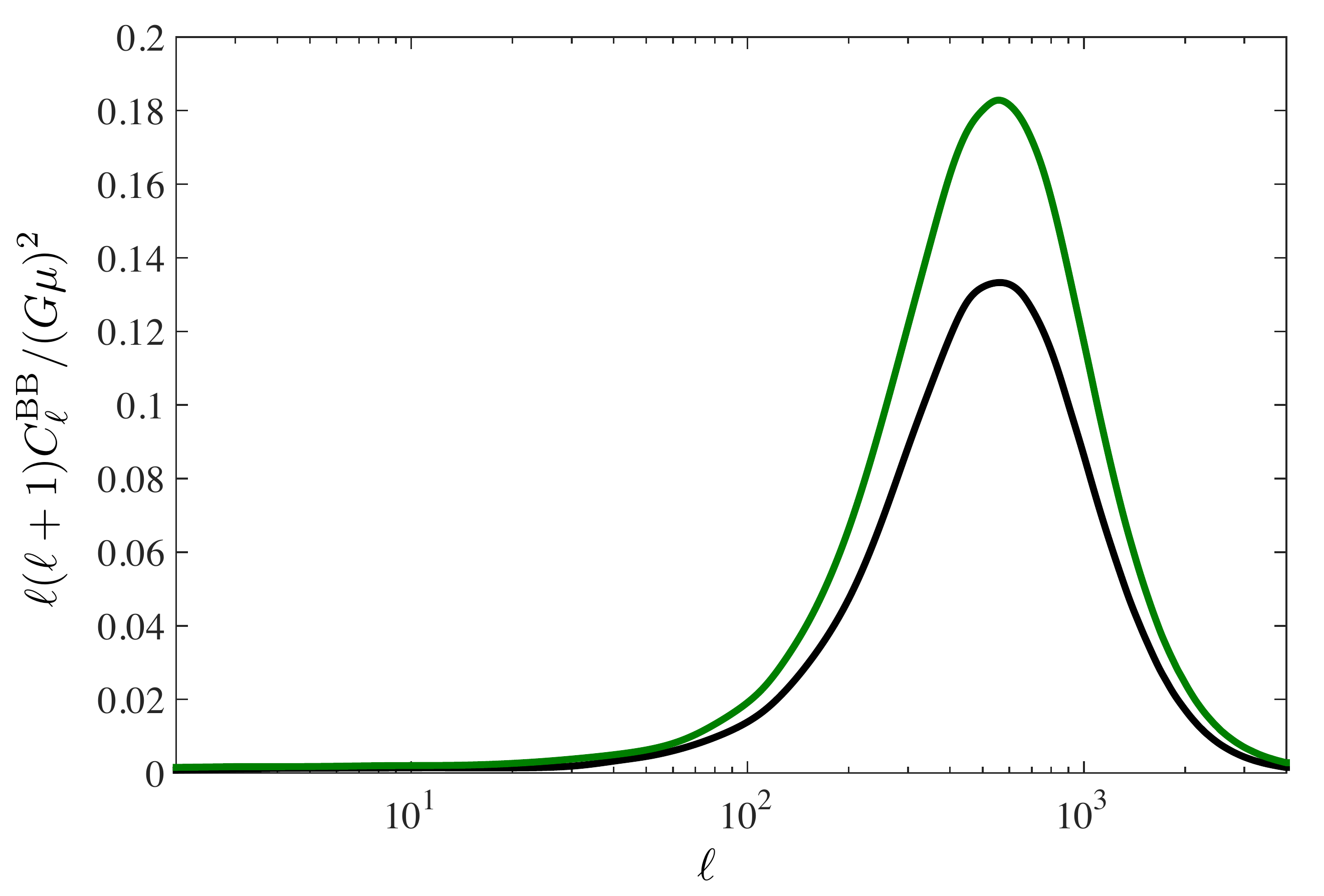}
\caption{Comparison of spectra obtained using same UETC data (based on \cite{Daverio:2015nva}) but different eigenvector interpolation methods. The black line represents the simple eigenvector decomposition (case 1) and the green line is for the multi-stage interpolation method (case 2). The main reason for the higher amplitude of the latter $C_\ell$ is the slower radiation-matter transition function.}
\label{figure_cl2}
\end{figure}

We show the comparison for this case in Figure~\ref{figure_cl2}. The spectra obtained by the simple eigenvector interpolation method is represented in black ({case 1}) and the multi-stage case in green ({case 2}). As it can be seen, in contrast to the previous case, the differences are substantial. The amplitude of the spectra, both in temperature and in B-mode channels, is considerably increased by the new transition treatment. The matter-radiation transition depicted by the ETCs and described by $f(\tau)$ is much slower than previous transition functions. The slower transition reflects in more contribution to the total from radiation correlators, which are higher in amplitude than the matter ones.

Table~\ref{table_percentualCl} shows the percentage changes of the height of the peak for the different UETC interpolation methods with respect to the simple eigenvector interpolation. It reflects that the increase affects all channels, but more considerably the polarization channels, which is a signal that the slower transition affects the vector and tensor perturbations more than the scalars.

\begin{table}[t]
\begin{center}
\renewcommand{\arraystretch}{1.2}
\begin{tabular}{|c|c|c|c|}
\hline
 & TT & EE & BB \\ \hline
Multistage & $+25\%$ & $+37\%$  & $+37\%$ \\ \hline
Fixed-$k$ & $+29\%$ & $+37\%$  & $+36\%$ \\
 \hline
\end{tabular}
 \caption{\label{table_percentualCl} Percentage changes of the height of the peak of CMB power spectra in TT, EE and BB calculated using the multistage (Sec.~\ref{subsec:multistage}) and fixed-$k$ (Sec.~\ref{subsec:fixedk}) interpolation methods with respect to the simple eigenvector interpolation method (Sec.~\ref{subsec:eigen}).}
\end{center}
\end{table}

%=====================================================================

\subsection{Case 2 $\rightarrow$ 2.1: Effects of the matter-$\La$ transition function}
\label{subsec_C}

\begin{figure}[h!]
\centering
\includegraphics[width=0.5\textwidth]{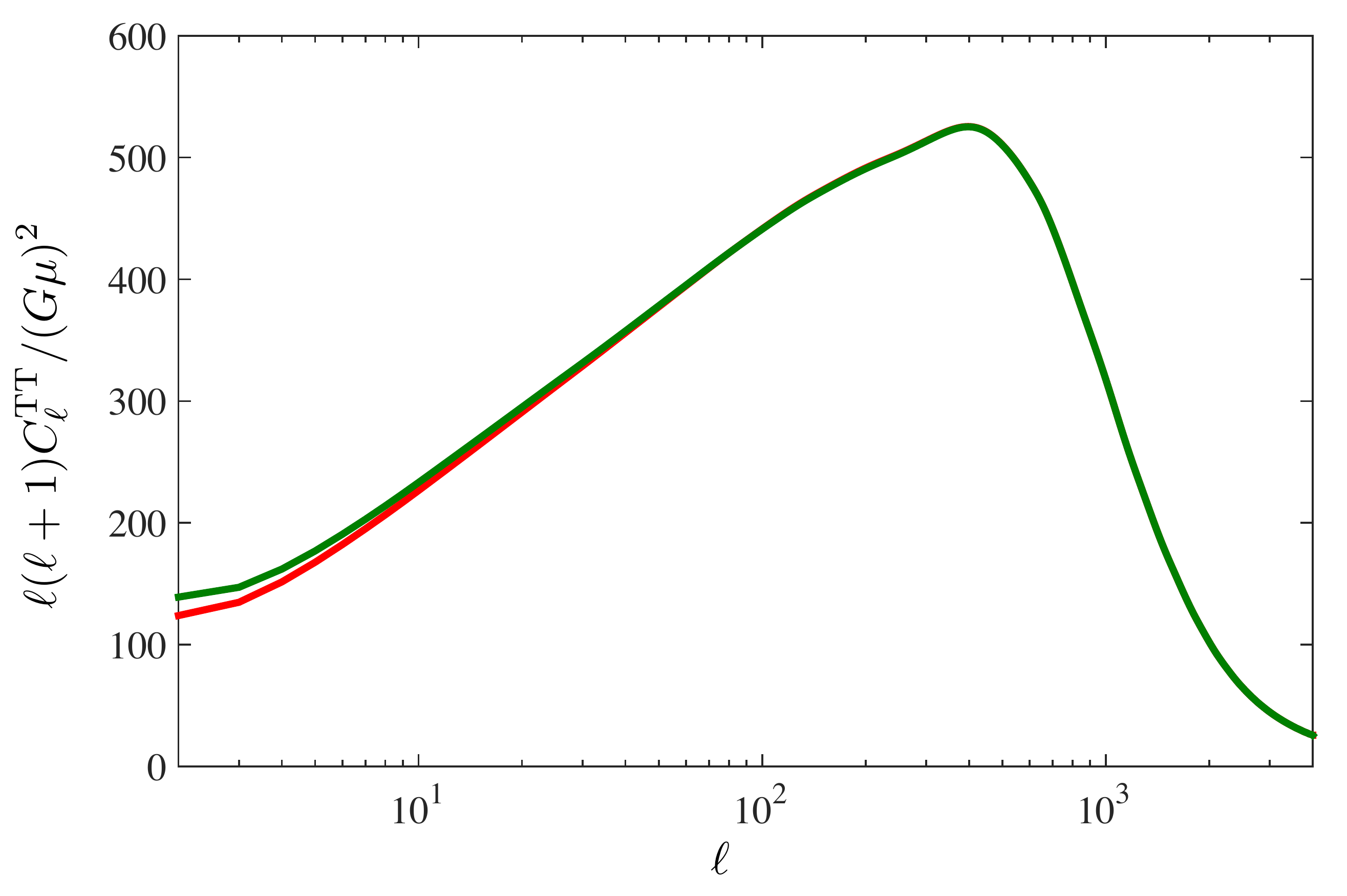}
\caption{Effect of including the matter-$\La$ transition into the multi-stage eigenvector interpolation structure, in green without the $\La$ transition and in red with it.}
\label{figure_cl3}
\end{figure}

The effect of including the matter-$\Lambda$ transition with its corresponding ETC-based interpolation function \cite{Daverio:2015nva} is represented in Figure~\ref{figure_cl3}, where the spectra that includes this latter transition is plotted in red ({case 2.1}) against the case without $\Lambda$ transition in green ({case 2}). In accordance with our previous expectations the inclusion of the matter-$\Lambda$ transition only affects scales that entered late into the horizon, \ie very low multipoles, decreasing slightly their amplitude.  

%=====================================================================

\subsection{Case 2 $\rightarrow$ 3: Effects of fixed-$k$ UETC interpolation}
\label{subsec_D}

Finally we calculate the spectra for the fixed-$k$ UETC interpolation method ({case 3}). This method interpolates UETCs in $k$-space rather than in time (see Eq.~(\ref{f:UETCmodel}) and similarly for the matter-$\Lambda$ transition). Conceptually this method fits naturally into usual Einstein-Boltzmann integrators, preserves the orthonormality of the source functions all the time and, as it was shown in \cite{Daverio:2015nva}, reproduces better the UETCs at the transitions. 

\begin{figure}[h!]
\centering
\includegraphics[width=0.49\textwidth]{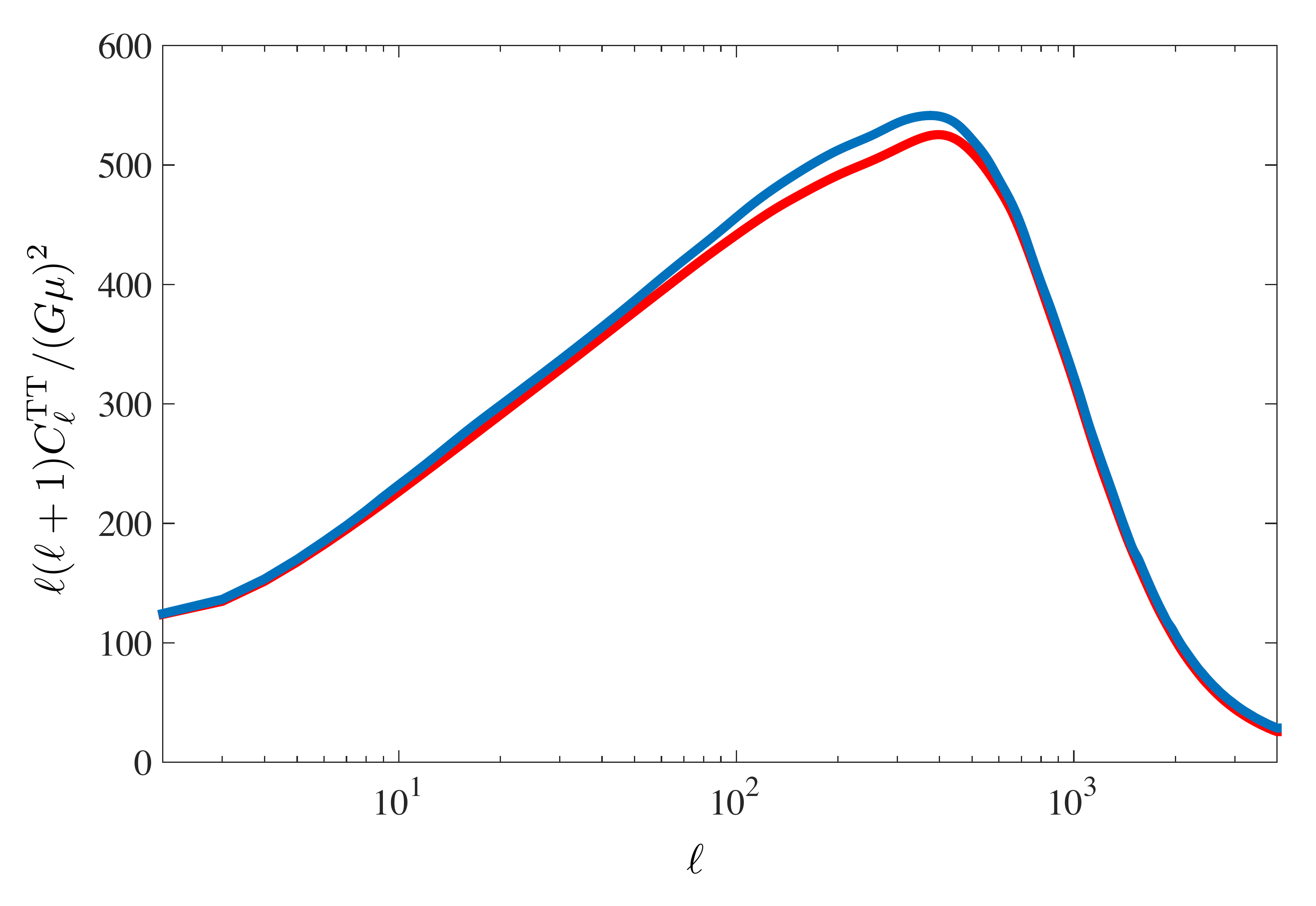}
\includegraphics[width=0.49\textwidth]{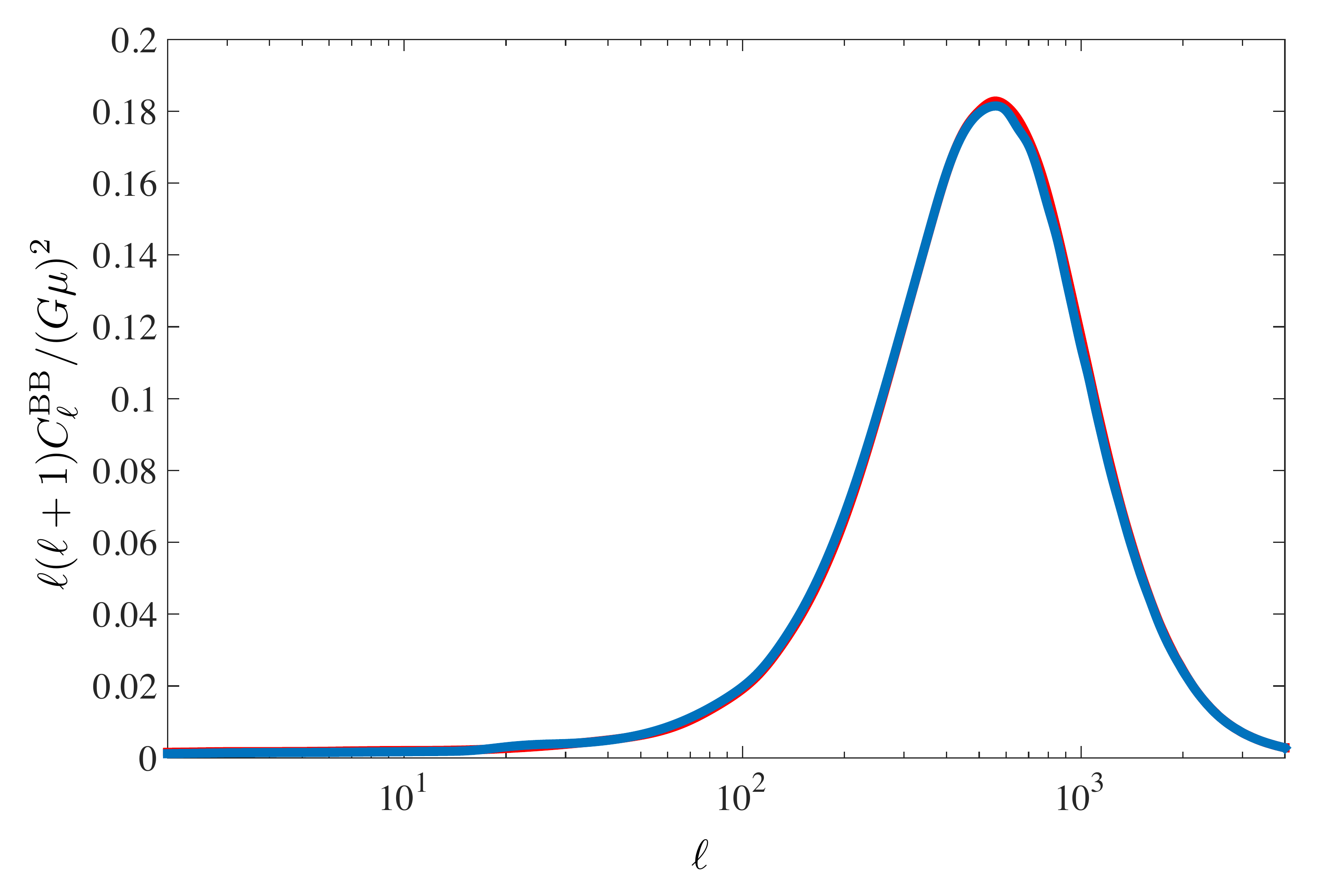}
\caption{Comparison of spectra obtained using different cosmological transition interpolation methods: multi-stage eigenvector interpolation (red line) and fixed-$k$ UETC interpolation (blue line).}
\label{figure_cl4}
\end{figure}

The effect of switching from the multi-stage interpolation method to the fixed-$k$ interpolation is shown in Figure~\ref{figure_cl4}, where the spectra of the multi-stage eigenvector set is plotted in red ({case 2.1}) and the result of the source function of the fixed-$k$ interpolated UETCs in blue ({case 3}). The matter-$\Lambda$ interpolation has also been included in both cases. By inspection we get that the minimum number of $k$ intervals needed to reach convergence in spectra is $N_{k}=55$. 

Figure~\ref{figure_cl4} and Table~\ref{table_percentualCl} show that the change introduced by the new interpolation method is small, only the amplitude of the spectra around the peak is slightly affected.

%=====================================================================

\subsection{Final $C_{\ell}$'s}

This section contains the final CMB anisotropy power spectra of this work. This new baseline set of $C_{\ell}$'s is based on the following ingredients:
\begin{enumerate}
	\item UETCs from $s=1$ and $s=0$ simulations on $4096^3$ sized grids and combined by the new UETC merging scheme.
	\item Cosmological transitions (radiation-matter and matter-$\Lambda$) modeled by fixed-$k$ UETC interpolation.
	\item Interpolation functions for radiation-matter and matter-$\Lambda$  transitions as described in Section \ref{subsec:fixedk}.
\end{enumerate}

\begin{figure}[h!]
\centering
\includegraphics[width=0.5\textwidth]{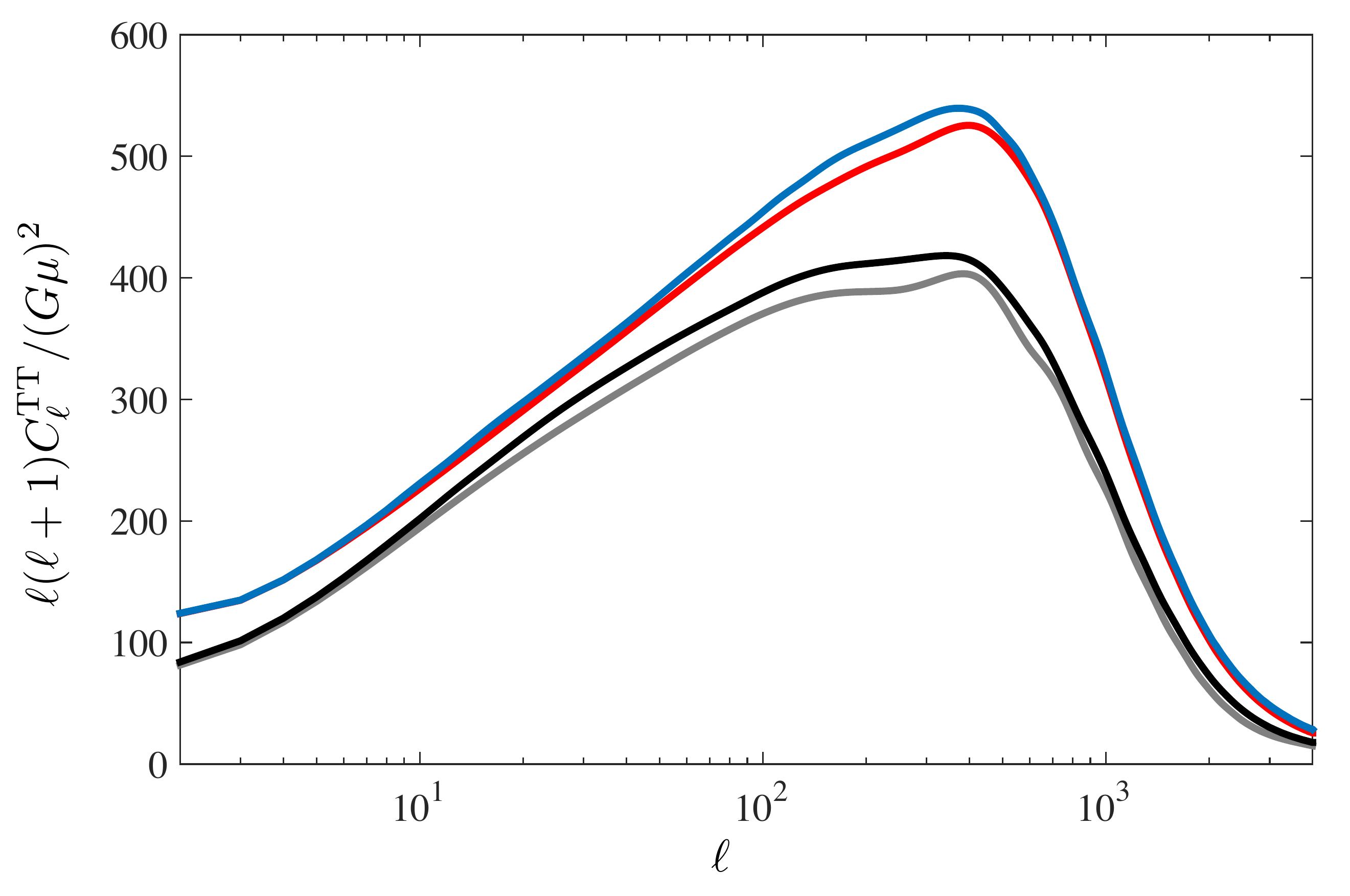}
\caption{Evolution of the power spectra produced by the improvements of this work. The baseline is the temperature angular power spectrum of Ref.~\cite{Bevis:2010gj} (grey). Also shown are $C_\ell^{TT}$ computed with same method (simple eigenvector interpolation) using the new UETCs (black). The effect of different treatment of the cosmological transitions can be seen in the spectra produced from the new UETCs by multi-stage eigenvector interpolation (red) and fixed-$k$ UETC interpolation (blue).}
\label{figure_cl5}
\end{figure}

The evolution of the temperature spectrum by the inclusion of new improvements is shown in Figure~\ref{figure_cl5}. We maintain the same color scheme as in previous plots: the spectrum of \cite{Bevis:2010gj} is the grey line, while the spectra obtained using new UETCs are shown in black (single eigenvector interpolation), red (multi-stage eigenvector interpolation) and blue (fixed-$k$ interpolation). The upward trend is clear and as we mentioned the main change is due to the new and slower radiation-matter interpolation function.

\begin{figure*}[tbh]
\centering
\includegraphics[width=0.95\textwidth]{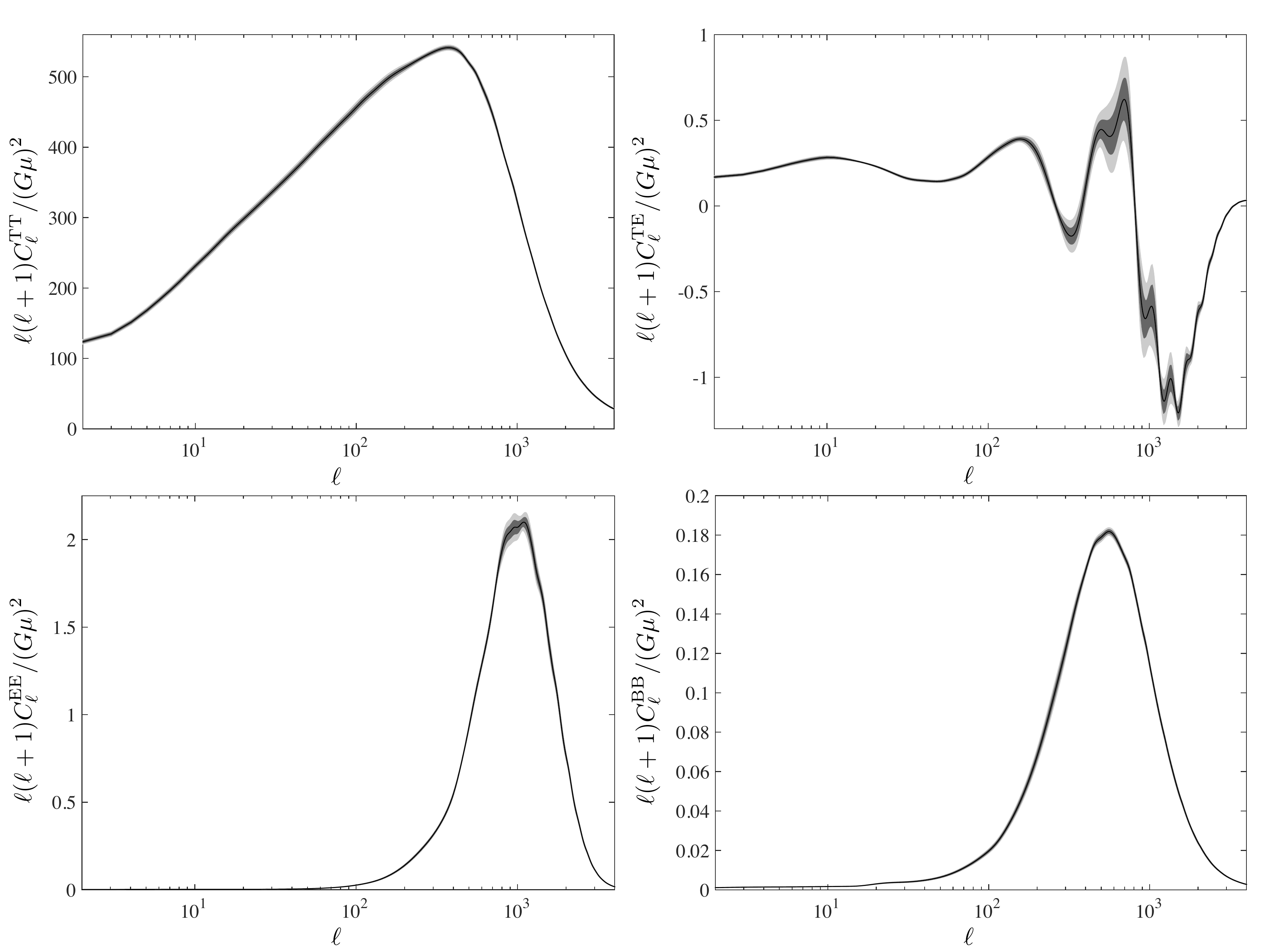}
\caption{Temperature and all polarization channels for the final CMB power spectra with all improvements implemented: new UETCs, fixed-$k$ UETC interpolation and new radiation-matter and matter-$\La$ transition functions. Black lines correspond to the mean spectra while grey regions represent $1\sigma$ and $2\sigma$ confidence limits obtained by bootstrapping 10 times over 7 radiation and 7 matter realizations in the UETC merging process.}
\label{figure_finalcl}
\end{figure*}

Our final power spectra with their corresponding error bars are included in Figure~\ref{figure_finalcl}. The black line represents the mean curve and the grey regions systematic errors obtained by bootstrapping 10 times over 7 radiation and 7 matter realizations in the UETC merging process.

%=====================================================================
%=====================================================================
%=====================================================================

\section{Fits and constraints}
\label{sec:fits}

We use the \Planck\ 2015 TT, TE, EE + lowTEB likelihood \cite{Aghanim:2015xee} as also used by the \Planck\ collaboration in \cite{Ade:2015xua} to put limits on the allowed fraction of cosmic strings and on other cosmological parameters. We always vary the standard six $\Lambda$CDM parameters of the basic inflationary model or ``power-law" (\PL) model, $\{\omega_b, \omega_c, \theta_{\rm MC}, \tau, \ln 10^{10} A_s, n_s\}$ and the nuisance parameters included in the likelihood, in addition to the tensor to scalar ratio $r$ and/or the string amplitude $10^{12} (G\mu)^2$. We use flat priors for all parameters that are wide enough to encompass the posteriors, with the conditions $r\geq0$ and $G\mu\geq0$. The primordial spectra are computed with the help of {\sc camb} \cite{Lewis:1999bs}, while we use {\sc cosmomc} \cite{Lewis:2002ah} for the Markov-Chain Monte Carlo (MCMC) sampler and to analyze the output. For the cosmic string spectra, we use the fixed-$k$ interpolation and vary only the string amplitude, \ie we keep the shape of the string $C_\ell$ fixed, as strings contribute at most of the order of 1\% to the total TT power on large angular scales. To quantify this, we introduce the parameter $f_{10}$ which is the ratio of the defect $C_\ell^{(TT)}$ to the total $C_\ell^{(TT)}$ at $\ell=10$.

The resulting 95\% upper limits on the string tension, $f_{10}$ and $r$ are given in table \ref{t:gmulimits}. The fit to the data is not improved either by adding primordial gravitational waves or cosmic strings, as can be seen from the constant value of $\mathcal{L}_{\rm max}$, and we find no significant preference for non-zero $r$ or $G\mu$. The results that we find for $G\mu=0$ are consistent with those of the \Planck\ analyses \cite{Ade:2015xua,Ade:2015lrj}. There are no significant degeneracies with any of the parameters varied here which has been the case since \cite{Urrestilla:2011gr} (but see also \cite{Lizarraga:2012mq} for degeneracies with other parameters not varied here). 

The constraints on $f_{10}$ that we find in the \PLgmu\ model are 13\% lower than those found by the \Planck\ collaboration, while the limits shown here on $(G\mu)^2$ are 30\% stronger. The former is due to the relatively small change in shape, while the stronger limit on $G\mu$ comes from the increase in amplitude (see Table \ref{table_percentualCl}), which is mostly due to the improved radiation-matter transition modelling.

\begin{table}
\begin{center}
\renewcommand{\arraystretch}{1.2}
\begin{tabular}{|c||c|c|c|c|} 
\hline
Dataset &  \multicolumn{4}{c|}{\Planck\ 2015 TT, TE, EE + low TEB} \\
\hline
Model & {\PLgmu} & {\PLrgmu} & \PLr &  \PL \\ \hline
$r$ & $-$ & $<0.11$ & $<0.11$ & $-$ \\ 
$f_{10}$ & $<0.013$  & $<0.011$  & $-$   & $-$   \\
$10^{12}(G\mu)^2$ & $<0.040$  & $<0.034$ & $-$ & $-$\\ 
\hline 
$-\ln{\cal L}_\mathrm{max}$ & $6472$ & $6472$ & $6472$ & $6472$ \\\hline
 \end{tabular} \\ 
  \caption{\label{t:gmulimits} 95\% upper limits for $(G\mu)^2$ and $r$ as well as best-fit likelihood values for different cosmological models, fitting for the \Planck\ 2015  TT, TE, EE and low TEB data. }
\end{center} 
\end{table}

%=====================================================================
%=====================================================================
%=====================================================================
\section{Discussion and conclusions}

We have calculated the CMB power spectra from the energy-momentum correlations computed in recent large-scale numerical simulations of a network of cosmic strings in the Abelian Higgs model \cite{Daverio:2015nva}, and compared them to \Planck\ CMB power spectra. We obtain a revised constraint on the cosmic string tension parameter $G\mu$, and investigate the source of the difference.

The new numerical simulations represent a significant improvement over those \cite{Bevis:2010gj} used in previous comparisons to \Planck\ data \cite{Ade:2015xua}, with a factor 64 increase in volume and sufficient resources to properly solve the equations in an expanding universe, and to investigate the radiation-matter and matter-$\La$ transitions, both for the first time.

The larger simulations confirmed the shape and normalisation of the unequal time correlators (UETCs) computed in \cite{Bevis:2010gj}, within our original uncertainty estimates. The biggest change comes from the improved method of treating the radiation-matter transition \cite{Daverio:2015nva}, which demonstrates that the larger UETCs of the radiation era are preserved for longer than previously thought. The consequence is that strings produce 30\% higher CMB power spectra for a given $G\mu$.

We trace the effect of the changes in Fig.~\ref{figure_cl5}. It can be seen that the TT power spectrum from the new simulations computed by the 2010 methods  are very close to the 2010 power spectra, and that the major change comes from improvements in the treatment of the transition in the computation. We investigate two new methods for CMB power spectrum computation from defects, one proposed in \cite{Fenu:2013tea}, and the other our own \cite{Daverio:2015nva}, which is conceptually simpler and easier to implement. They differ little: the main effect comes from the increase in amplitude of the UETCs at the time of decoupling, when the CMB perturbations are generated.

Given the agreement between the old and new simulations and the new treatment of the cosmological transitions, we expect that the main theoretical uncertainty in the resulting $C_\ell$ is now due to the old Einstein-Boltzmann solver that we use to compute the power spectrum from the UETC. That code is based on an old version of CMBEasy \cite{Doran:2003sy} with a by-now outdated recombination history and other legacy precision issues. In the future we plan to move to CLASS \cite{Lesgourgues:2011re}, a project that is currently under way.

Our final constraints are given in Table \ref{t:gmulimits}, quoted as limits on the fraction of the power spectrum due to cosmic strings at multipole $\ell=10$ $\fd$, and the square of the string tension parameter $(G\mu)^2$.  We see that whether or not we allow for primordial tensor fluctuations in the model, no more than about 1\% of the CMB power spectrum can be due to cosmic strings.  With our improved calculation of the power spectrum, the limit on $G\mu$ (assuming no primordial tensor fluctuations) is approximately $2.0\times 10^{-7}$, 17\% lower than the value $2.4\times10^{-7}$ obtained with the old method \cite{Ade:2015xua}. The new constraint corresponds to a symmetry-breaking scale of $2.2 \times 10^{15}$ GeV in the Abelian Higgs model at critical coupling.

%================================================================================
%================================================================================
%================================================================================

\vspace{1cm}
\begin{acknowledgments}
We thank Neil Bevis and Ruth Durrer for helpful discussions. This work has been supported by two grants from the Swiss National Supercomputing Centre (CSCS) under project IDs s319 and s546. In addition  this work has been possible thanks to the computing infrastructure of the i2Basque academic network, the COSMOS Consortium supercomputer (within the DiRAC Facility jointly funded by STFC and the Large Facilities Capital Fund of BIS), and the Andromeda/Baobab cluster of the University of Geneva. JL and JU acknowledge support from Ministerio de Econom\'\i a y Competitividad (FPA2015-64041-C2-1-P) and  Consolider-Ingenio Programme EPI (CSD2010-00064). DD and MK acknowledge financial support from the Swiss NSF. MH acknowledges support from the Science and Technology Facilities Council (grant number ST/L000504/1).
\end{acknowledgments}

\vspace{3mm}

\bibliographystyle{h-physrev4}% This is very necessary for compilation of the BIbtex sources, why!?
\bibliography{CosmicStrings.bib}

\end{document}